\documentclass{jfm}
\usepackage{graphicx}
\usepackage{psfrag}
\usepackage{natbib}
\usepackage{amsmath}
\usepackage{amssymb}
\usepackage{wasysym}

\title{Gravity currents in a porous medium at an inclined plane}
\author[D. Vella \& H. E. Huppert]{D\ls O\ls M\ls I\ls N\ls I\ls C\ns
  V\ls E\ls L\ls L\ls A \and H\ls E\ls R\ls B\ls E\ls R\ls T\ns E.\ns
  H\ls U\ls P\ls P\ls E\ls R\ls T}
\affiliation{Institute of Theoretical Geophysics, Department of Applied Mathematics and Theoretical Physics, University of Cambridge, Wilberforce Road, Cambridge, CB3 0WA, U.\ K.}
\begin{document}

\maketitle

\begin{abstract}
We consider the release from a point source of relatively heavy fluid
into a porous saturated medium above an impermeable slope. We consider
the case where the volume of the resulting gravity current increases
with time like $t^\alpha$ and show that for $\alpha<3$, at short times the current
spreads axisymmetrically, with radius $r\sim t^{(\alpha+1)/4}$, while
at long times it spreads
predominantly downslope. In particular, for long times the downslope position of the
current scales like $t$ while the current extends a distance
$t^{\alpha/3}$ across the slope. For $\alpha>3$, this situation is
reversed with spreading occurring predominantly downslope for short
times. The governing equations admit similarity solutions whose scaling behaviour we
determine, with  the full similarity form being evaluated by numerical computations of the governing
partial differential equation. We find that the results of these
analyses are  in good quantitative agreement with a series of
laboratory experiments. Finally, we discuss the implications
of our work for the sequestration of carbon dioxide in aquifers with a
sloping, impermeable cap.
\end{abstract}

\section{Introduction}

Horizontal differences in density between two fluids lead to the
propagation of so-called gravity currents. These currents are of
interest in a number of industrial as well as natural applications
and so obtaining an understanding of the way in which they propagate
is a subject that has motivated a considerable amount of current
research~\cite[][]{huppert06}. 

In previous publications, our understanding of axisymmetric viscous gravity
currents on an impermeable boundary~\cite[][]{huppert82a} has been generalised to take
account of the effects of a slope~\cite[][]{lister92} as well as the
propagation of a current in a porous
medium~\cite[][]{huppert95,lyle05}. Here, we consider the propagation of a
gravity current from a point source in a porous
medium at an impermeable sloping boundary. Of particular interest is
the evolution of the current away from the axisymmetric similarity solution found
by \cite{lyle05}.

We begin by deriving the evolution equations for the
shape of a current whose volume varies in time like
$qt^{\alpha}$. A scaling analysis of these governing equations reveals the extent of the current as a function of time up to a
multiplicative constant. The full form of the similarity solutions
that give rise to these scalings can only be determined by numerical
means, however, and to do so we modify the numerical code
of~\cite{lister92}. For some particular values of $\alpha$,
it is possible to make analytical progress; these cases are considered
separately and provide a useful check of the numerical scheme. We then
compare the results of the numerical calculations to a series of
experiments and find good quantitative agreement between the
two. Finally, in the last section, we discuss the implications of our results in geological
settings, with particular emphasis on the implications of our work for the
sequestration of carbon dioxide.

\begin{figure}
\centering
\includegraphics[height=7cm]{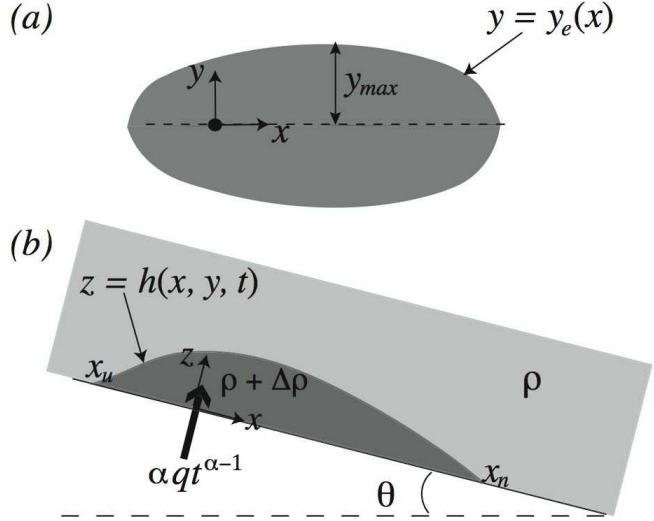}
\caption{Sketches of a gravity current, of density $\rho+\Delta\rho$, propagating
  in a porous medium saturated with liquid of density $\rho$ above an inclined plane. (a) Plan view of the current and (b) horizontal section through the current. }
\label{setup}
\end{figure}

\section{Formulation}

\subsection{Governing equations}

We consider a gravity current consisting of fluid material of density
$\rho+\Delta\rho$ in a deep porous medium saturated with fluid of density
$\rho$, which is bounded by an impermeable barrier at an angle
$\theta$ to the horizontal (see figure~\ref{setup} for a sketch of the
setup). That the saturated porous medium is deep in comparison with the vertical extent of the current allows us to neglect the motion of the surrounding fluid, simplifying the problem considerably. We use the natural Cartesian co-ordinate system centred on the
mass source and aligned with the slope of the impermeable
boundary. The depth, $h(x,y,t)$, of the gravity current is then
determined by continuity combined with Darcy's law~\cite[see][for
  example]{bear88} and the assumption that the pressure in the current is hydrostatic, i.e.
\begin{equation}
P-P_0=\Delta\rho gh\cos\theta-(\rho+\Delta\rho)gz\cos\theta+\rho gx\sin\theta\hspace{0.5cm}(z<h),
\end{equation} with $P_0$ constant. Here, Darcy's law takes the form
\begin{equation}
\mbox{\boldmath{$u$}}=-\frac{k}{\mu}\left[\mbox{\boldmath{$\nabla$}}P-(\rho+\Delta\rho)g\bigl(\sin\theta,0,-\cos\theta\bigr)\right],
\end{equation} where $k$ is the permeability of the porous medium and
$\mu$ is the viscosity of the liquid. The velocity within the porous medium is therefore given by
\begin{equation}
\mbox{\boldmath{$u$}}=-\frac{k\Delta\rho g}{\mu}\left(-\sin\theta+\cos\theta\frac{\partial h}{\partial x},\cos\theta\frac{\partial h}{\partial y},0\right).
\label{darcyvel}
\end{equation}  Using this along with the
conservation of mass, we obtain
\begin{equation}
\frac{\partial h}{\partial t}=\frac{k\rho g'}{\mu\phi}\left(\frac{\cos\theta}{2}\nabla^2h^2-\sin\theta\frac{\partial h}{\partial x}\right),
\label{pde}
\end{equation} where $\phi$ is the porosity of the porous medium and
$g'\equiv g\Delta\rho/\rho$. Equation \eqref{pde} is a nonlinear
advection--diffusion equation for the current thickness, with the two
terms on the right hand side representing the gravity--driven
spreading of the current and its advection downslope, respectively.

It is common to close the system by requiring that the volume of the
current depend on time like $qt^\alpha$ for some constant
$\alpha\geq0$~\cite[][]{huppert82a,lister92,huppert95}. This constraint
leads to solutions of self-similar form (as we shall see again in this
case) but also covers the natural cases of a fixed volume release ($\alpha=0$) and a constant flux release ($\alpha=1$). To impose this volume constraint, (\ref{pde}) must be solved
along with
\begin{equation}
\phi\int_{x_u}^{x_n}\int_{-y_e(x)}^{y_e(x)}h\mathrm{~d}y\mathrm{~d}x=qt^\alpha,
\label{vol_const}
\end{equation} with $|y|=y_e(x)$ giving the edge of the current for
$x_u(t)<x< x_n(t)$. Note that \eqref{vol_const} contains an extra multiplicative factor of $\phi$, which was  omitted in the study of an axisymmetric current in a porous medium by \cite{lyle05}.

Equations (\ref{pde}) and (\ref{vol_const}) may be non-dimensionalized by setting $T=t/t^*$, $H=h/h^*$, $X=x/x^*$ and $Y=y/y^*$, where
\begin{equation}
t^*\equiv
\left(\frac{q}{\phi V^3\tan\theta}\right)^{\frac{1}{3-\alpha}},\hspace{0.5cm}
x^*=y^*\equiv Vt^*,
\hspace{0.5cm}h^*\equiv x^*\tan\theta,
\label{scales}
\end{equation} and
\begin{equation}
V\equiv\frac{k\rho g'\sin\theta}{\mu\phi}
\label{natvel}
\end{equation} is the natural velocity scale in the problem. In non--dimensional terms, therefore, the current satisfies
\begin{equation}
\frac{\partial H}{\partial
 T}=\mbox{\boldmath{$\nabla$}}\cdot(H\mbox{\boldmath{$\nabla$}}
 H)-\frac{\partial H}{\partial X},
\label{ndpde}
\end{equation} along with the volume conservation constraint
\begin{equation}
 \int_{X_u}^{X_n}\int_{-Y_e(X)}^{Y_e(X)}H\mathrm{~d}Y\mathrm{~d}X=T^\alpha.
\label{ndvol}
\end{equation}

\subsection{Scalings\label{secscalings}}

To aid our physical understanding of the spreading of the gravity
current, we begin by considering the scaling behaviour of the
spreading in the limits of short and long times. For $\alpha<3$, \eqref{darcyvel} shows that at short times ($T\ll 1$) the typical horizontal velocity scale is $X/T\sim H/X$ so that $H\sim X^2/T$. Further, $X\sim Y$ and volume conservation \eqref{ndvol} requires that $HXY\sim T^\alpha$. From this we therefore find  the axisymmetric scalings obtained by~\cite{lyle05}, namely
\begin{equation}
H\sim T^{\frac{\alpha-1}{2}},\hspace{0.5cm} X\sim Y\sim
T^{\frac{\alpha+1}{4}}.
\label{axiscale}
\end{equation} At long times ($T\gg1$), again for $\alpha<3$, the typical downslope velocity of the current is $X/T\sim 1$ while in the across-slope direction we have $Y/T\sim H/Y$. Combined with volume conservation $XYH\sim T^\alpha$ these scalings lead to
\begin{equation}
H\sim T^{\frac{2\alpha-3}{3}},\hspace{0.5cm} X\sim T,\hspace{0.5cm}Y\sim T^{\frac{\alpha}{3}},
\label{scalings}
\end{equation} so that the current spreads predominantly downslope. It is worth noting here that the long time scaling $X\sim T$ is
unsurprising because \eqref{ndpde} may be simplified by moving into a
frame moving at unit speed downslope~\cite[][]{huppert95}. We also note that the scaling
$Y\sim T^{\alpha/3}$ is identical to that found by~\cite{lister92} for a
viscous current on a slope.

When $\alpha>3$, the importance of the two downslope terms
(the diffusive and translational terms) reverses. In particular, at long
times $(HH_X)_X\gg H_X$, so that we in fact recover the
axisymmetric spreading scalings given in \eqref{axiscale} as being relevant
for $T\gg1$. Conversely, for $T\ll1$ we recover the non-axisymmetric
scalings of \eqref{scalings}. A summary of the different scaling regimes expected is given in dimensional terms in table \ref{tabl_scale}.

That we observe axisymmetric spreading if $\alpha>3$ and $T\gg1$ is surprising,
but is a consequence of the fact that the downslope flux in a porous
medium gravity current is only
weakly dependent on the local height and so can be
swamped by the spreading terms in \eqref{ndpde}. In the viscous case,
this is not possible because the downslope flux is able to remove the
incoming flux much more efficiently and penalizes the accumulation of
material at a particular point more.

\begin{table} 
\centering
\setlength{\tabcolsep}{0.6 em}
\begin{tabular}{ccccc}
 \multicolumn{2}{c}{Regime} & Downslope extent & Cross-slope extent & Thickness \\
& & $x$ & $y$ &$h$\\
$\alpha<3$ & $t\ll t^*$ & $\sim \left(\frac{Vq}{\phi\tan\theta}\right)^{1/4}t^{(\alpha+1)/4}$ &$\sim \left(\frac{Vq}{\phi\tan\theta}\right)^{1/4}t^{(\alpha+1)/4}$& $\sim\left(\frac{q\tan\theta}{\phi V}\right)^{1/2}t^{(\alpha-1)/2}$\\
$\alpha<3$ & $t\gg t^*$ & $\sim V t$ &$\sim\left(\frac{q}{\phi \tan\theta}\right)^{1/3}t^{\alpha/3}$& $\sim \left(\frac{q^2\tan\theta}{\phi^2V^3}\right)^{1/3}t^{(2\alpha-3)/3}$\\
$\alpha>3$ & $t\ll t^*$ & $\sim V t$ &$\sim\left(\frac{q}{\phi \tan\theta}\right)^{1/3}t^{\alpha/3}$& $\sim \left(\frac{q^2\tan\theta}{\phi^2V^3}\right)^{1/3}t^{(2\alpha-3)/3}$\\
$\alpha>3$ & $t\gg t^*$ & $\sim \left(\frac{Vq}{\phi\tan\theta}\right)^{1/4}t^{(\alpha+1)/4}$ &$\sim \left(\frac{Vq}{\phi\tan\theta}\right)^{1/4}t^{(\alpha+1)/4}$& $\sim\left(\frac{q\tan\theta}{\phi V}\right)^{1/2}t^{(\alpha-1)/2}$\\
\end{tabular}
\caption{Summary of the asymptotic scalings for the dimensions of a
  gravity current in a porous medium at an inclined plane. Here
  dimensional notation is used for clarity, and $t^*$ and $V$ are as
  defined in \eqref{scales} and \eqref{natvel}, respectively.}
\label{tabl_scale}
\end{table}

\subsection{\label{numerics}Numerics}

The axisymmetric spreading of a gravity current in a porous medium
above an horizontal plane was considered by~\cite{lyle05}. In
particular, they determined the coefficients in the scalings
\eqref{axiscale} by finding a solution dependent on one similarity variable in this case. To determine the prefactors in the non-axisymmetric scaling relations \eqref{scalings},
it is necessary to resort to numerical solutions of \eqref{ndpde} and
\eqref{ndvol}. The numerical code used to do this was adapted from that used by~\cite{lister92}
for a viscous gravity current on an inclined plane, with minor
alterations to make it applicable to a gravity current in a porous
medium. This code is an implementation of a finite-difference scheme on a rectangular grid
with time-stepping performed using an alternating-direction-implicit
method. Equation \eqref{ndpde} was written in flux-conservative form
allowing the diffusive and advective terms to be represented by the
Il'in scheme~\cite[][]{clauser87}. More details of the numerical
scheme may be found in~\cite{lister92}.

\section{Special values of $\alpha$}

In this section, we consider separately particular values of
$\alpha$ that are of special interest. In some of these cases, it is possible to make progress analytically providing useful checks on the numerical scheme discussed in
section \ref{numerics}, but they also shed light on situations of
practical interest.

\subsection{Constant volume}

As already noted, the differential equation in \eqref{ndpde} may be simplified by moving
into a frame translating at unit speed downslope. However, for
general values of $\alpha$, this corresponds to a point source that
is moving uphill in the new frame, complicating the analysis. For a current of constant volume,
$\alpha=0$, there is no distinguished source point and we let
$X'\equiv X-T$. The resulting transformation of \eqref{ndpde} has an
axisymmetric similarity solution~\cite[][]{lyle05}, which may be written
\begin{equation}
H(X,Y,T)=\frac{1}{8T^{1/2}}\left(\frac{4}{\sqrt{\pi}}-\frac{{R'}^2}{T^{1/2}}\right),
\end{equation} where $R'\equiv ({X'}^2+Y^2)^{1/2}$.

\subsection{Constant flux: A steady state}

For very long times $T\ggg1$, we expect that a constant flux current (corresponding to $\alpha=1$)
will approach a steady state, whose shape we now determine. We expect
this steady shape to be observed far from the nose of the current,
since the nose is always unsteady, requiring that $X\ll T$. Sufficiently far downstream from the source ($X\gg1$), the steady shape is given by
\begin{equation}
\frac{\partial^2 H^2}{\partial Y^2}=2\frac{\partial H}{\partial X},
\label{sseqn}
\end{equation} which has a similarity solution of the form $H(X,Y)=X^{-1/3}f(Y/X^{1/3})$ where the function $f$ satisfies
\begin{equation}
\frac{d^2f^2}{d\eta^2}+\frac{2}{3}\left(f+\eta \frac{df}{d\eta}\right)=0,\hspace{0.5cm} \int_{-\eta_e}^{\eta_e}f\mathrm{~d}\eta=1, \hspace{0.5cm} f(\pm\eta_e)=0.
\end{equation} This has solution
\begin{equation}
f(\eta)=\frac{1}{6}(\eta_e^2-\eta^2),
\label{simsoln}
\end{equation} where $\eta_e=(9/2)^{1/3}\approx1.651$ denotes the
position of the current edge in similarity variables. 

\begin{figure}
\centering
\includegraphics[height=7cm]{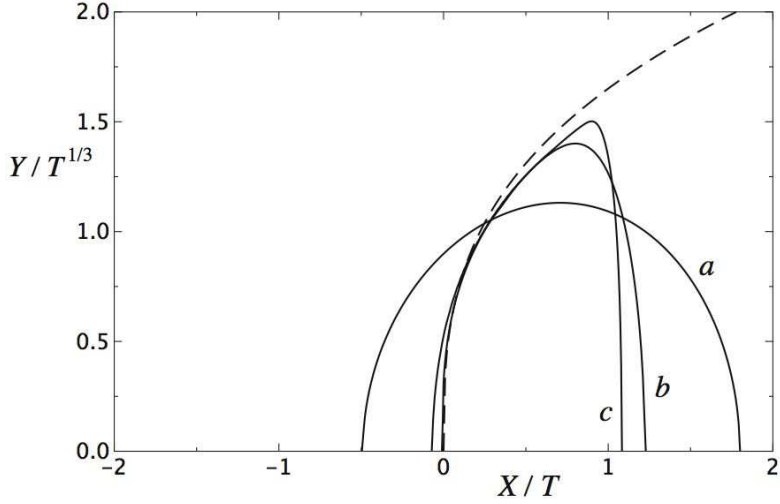}
\caption{Numerical evolution of the boundary of the current in rescaled
  co-ordinates at (a) $T=1.23$, (b) $T=9.52$ and (c) $T=270.9$. The
  last of these is indistinguishable from the steady state shape that
  is found at long times in these rescaled variables. The similarity
  solution for the steady shape in the interior is given by $Y=(9X/2)^{1/3}$ (dashed
  line) and is valid away from the source and the front regions, which in
  these rescaled variables requires that $T^{-1}\ll X/T\ll1$.}
\label{gc_num}
\end{figure}

This shows that far away from the source and nose regions, we should
expect the shape of unsteady currents to approach
$Y=(9X/2)^{1/3}$. Superimposing this curve onto the numerically
calculated current provides a useful check of the numerical scheme
described in section \ref{numerics}. This comparison (see figure~\ref{gc_num})
shows that, away from both the nose and source regions, we do indeed
see the steady state shape, though this region is confined to
$T^{-1}\ll X/T\ll1$ in the rescaled co-ordinates used in
figure~\ref{gc_num}.

It is interesting to note that the similarity solution \eqref{simsoln} is precisely that given by~\cite{huppert95} for the shape of a two-dimensional current of constant volume spreading in a porous medium above an horizontal boundary. This correspondence arises because in the steady state case considered here, fluid moves downslope at a constant velocity --- independently of its cross-slope position and the current height --- so that $X$ is a proxy for time. A material slice in the $y$--$z$ plane thus remains planar as it is advected downslope and so spreads laterally in exactly the same way that a fixed volume release does in two-dimensions.  

\subsection{$\alpha=3$}

When $\alpha=3$, the non-dimensionalization leading to \eqref{ndpde}
breaks down because there is no longer a characteristic time-scale
$t^*$ of the motion. Instead, an additional natural velocity
scale, $(q/\phi)^{1/3}$, enters the problem. We thus define a new set of dimensionless variables $\tilde{T}=t/\tilde{t}^*$, $\tilde{H}=h/\tilde{h}^*$, $\tilde{X}=x/\tilde{x}^*$ and $\tilde{Y}=y/\tilde{y}^*$ where $\tilde{t}^*$ is an arbitrary timescale and
\begin{equation}
\tilde{x}^*=\tilde{y}^*\equiv \left(\frac{q}{\phi\tan\theta}\right)^{1/3}\tilde{t}^*,\hspace{0.5cm} \tilde{h}^*\equiv\tilde{x}^*\tan\theta.
\end{equation} In these non-dimensional variables, the system becomes
\begin{equation}
\frac{\partial \tilde{H}}{\partial\tilde{T}}=\delta\left(\mbox{\boldmath{$\nabla$}}\cdot(\tilde{H}\mbox{\boldmath{$\nabla$}}\tilde{H})-\frac{\partial \tilde{H}}{\partial \tilde{X}}\right),
\label{a3:orpde}
\end{equation} along with volume conservation in the form
\begin{equation}
 \int_{\tilde{X}_u}^{\tilde{X}_n}\int_{-\tilde{Y}_e(\tilde{X})}^{\tilde{Y}_e(\tilde{X})}\tilde{H}\mathrm{~d}\tilde{Y}\mathrm{~d}\tilde{X}=\tilde{T}^3,
\label{a3:orvol}
\end{equation} where $\delta\equiv V(\phi\tan\theta/q)^{1/3}$ is essentially the ratio of the two velocity scales in the problem. By substituting $\tilde{H}=\tilde{T}\Phi(\xi,\eta)$ with $\tilde{X}=\tilde{T}\xi$ and $\tilde{Y}=\tilde{T}\eta$, time can be eliminated from this problem completely so that $\Phi$ is the solution of the two-dimensional problem
\begin{equation}
3\Phi=\left[\Phi\bigl(\xi+\delta(\Phi_\xi-1)\bigr)\right]_\xi+\left[\Phi(\delta\Phi_\eta+\eta)\right]_\eta,
\label{a3:pde}
\end{equation} (with subscripts denoting differentiation) and 
\begin{equation}
\int_{\xi_u}^{\xi_n}\int_{-\eta_e}^{\eta_e}\Phi\mathrm{~d}\eta\mathrm{~d}\xi=1.
\label{a3:vol}
\end{equation}

\begin{figure}
\centering
\includegraphics[height=5cm]{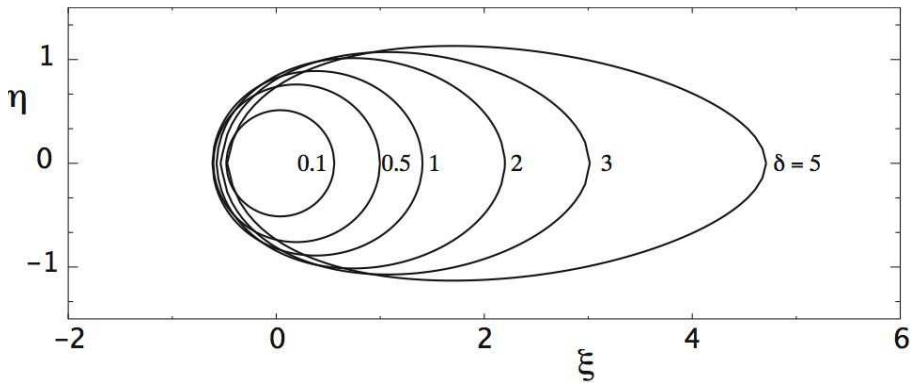}
\caption{Numerical results showing the shape of currents with $\alpha=3$ obtained by solving \eqref{a3:pde} and \eqref{a3:vol} for six values of the parameter $\delta$. Labels refer to the value of $\delta$ for each current.}
\label{a=3}
\end{figure}

The system \eqref{a3:pde} and \eqref{a3:vol} was solved by timestepping the problem in \eqref{a3:orpde} and \eqref{a3:orvol} using a minor modification of the code described in section \ref{numerics}. This was found to be a convenient method of solution and also demonstrates that time-dependent solutions  converge on the time-independent solution. The results of this calculation are shown in figure~\ref{a=3} for a number of different values of $\delta$.

The importance of the case $\alpha=3$ as a transition between
qualitatively different flow regimes is reminiscent of earlier work
on gravity currents. For an axisymmetric
 gravity current,~\cite{huppert82a} found that viscous forces
 dominate inertia at long times for $\alpha<3$ (being insignificant
 at short times) with the situation reversed for $\alpha>3$. \cite{acton01}
 found that a viscous gravity current propagating over a permeable
 medium spreads only a finite distance if $\alpha<3$ but spreads
 indefinitely for $\alpha>3$. Despite these similarities, the
 reappearance of a transition at $\alpha=3$ here is purely coincidental.

\subsection{$\alpha>3$}

\begin{figure}
\centering
\includegraphics[height=6cm]{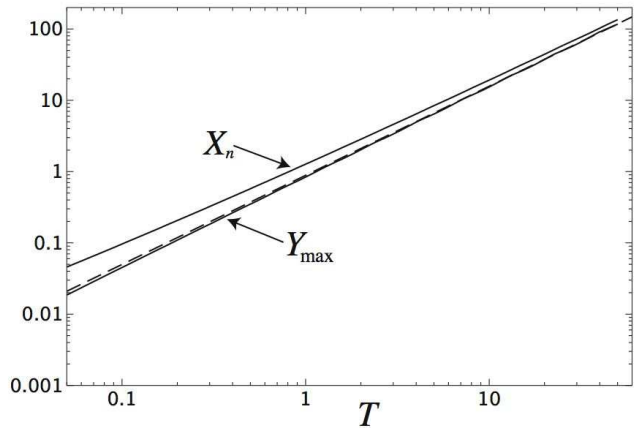}
\caption{Numerical results for the positions of the current edge $X_n$
  and $Y_{\mathrm{max}}$ as a function of time $T$ for $\alpha=4$
  (solid lines). For $T\gg1$ these obey the axisymmetric spreading
  relationship, $X_n, Y_{\mathrm{max}}\sim0.8855T^{5/4}$ (dashed
  line), that we expect from the axisymmetric analysis of~\cite{lyle05}.}
\label{alpha=4}
\end{figure}

In section \ref{secscalings}, we observed that for $\alpha>3$ a scaling analysis suggests that we should observe axisymmetric spreading for $T\gg1$. For such values of $\alpha$, therefore, we expect to recover the axisymmetric solutions given by~\cite{lyle05} in our numerical simulations. In particular, for $\alpha=4$ we would expect to find that
$$X_n, Y_{\mathrm{max}}\sim0.8855T^{5/4},$$ where the prefactor here has been determined by repeating the analysis of~\cite{lyle05}. As shown in figure~\ref{alpha=4}, this result is indeed obtained from our numerical results.

\section{Experimental results}
We conducted a series of experiments in which a saline solution (dyed red) was injected at
constant flux ($\alpha=1$) into the base of a porous medium saturated
with fresh water. The details of the experimental setup are as
described by \cite{lyle05}. In summary, the experiments were performed in a square-based Perspex tank of internal side length $61\mathrm{~cm}$ and height $41\mathrm{~cm}$. The porous medium consisted of a self-supported matrix of Glass ballotini (diameter $3\mathrm{~mm}$), which filled the tank to a height of $25\mathrm{~cm}$. In contrast to the experiments of~\cite{lyle05}, the Perspex tank was tilted (so that the gravity current was propagating on a slope) and the saline
solution was injected at the edge of the tank, away from the corner
because the inherent symmetry is different here to that of the axisymmetric
case. Video footage of the motion was
captured using a CCD camera and measurements of the front distance down slope $x_n$ as
well as the maximum lateral extent of the current $y_{\mathrm{max}}$
were made using the image analysis software ImageJ\footnote{ImageJ is
  distributed by the National Institutes of Health and may be
  downloaded from:
\texttt{http://rsb.info.nih.gov/ij/}}. The details of the six different values of $g'$, $q$ and $\theta$
investigated are given in table \ref{expt_dets}, along with the
relevant values of the typical scales $t^*$, $x^*$ and $h^*$. The
latter estimates are based on the measurements of $\phi=0.37$ and
$k=6.8\times10^{-9}\textrm{ m}^2$ given by \cite{lyle05}. The experimental results of \cite{lyle05} are in very good agreement with theory once the additional factor of $\phi$ in \eqref{vol_const} is included. We therefore believe these values of $\phi$ and $k$ to be correct.

\begin{table} 
\centering
\setlength{\tabcolsep}{0.9 em}
\begin{tabular}{cccccccc}
\emph{Expt.} & \emph{Symbol} & $g'$ ($\textrm{cm s}^{-2}$)  & $q$ ($\textrm{cm}^3 \textrm{s}^{-1}$) & $\theta$ ($^\circ$) & $t^*$ (s) &$x^*$ (m)  &$h^*$ (m) \\
$1$ & $\triangle$& $91$ & $2.14$ & $9.5$ &$40.5$&$0.112$&$0.019$ \\ 
 $2$ & $\Box$ &$99$ & $1.31$  & $10$&$25.2$&$0.080$&$0.014$\\ 
  $3$  & $\Diamond$& $99$ & $3.04$ &$18$&$11.9$&$0.067$&$0.022$ \\ 
 $4$ & \CIRCLE&$99$& $4$ &$18$ &$13.7$&$0.077$&$0.025$\\
  $5$ &$\blacksquare$& $99$ & $5.78$ &$18$ &$16.5$&$0.093$&$0.030$\\
  $6$ &$\bigstar$& $91$ & $3.86$ &$5$ &$196.2$&$0.286$&$0.025$\\
\end{tabular}
\caption{Parameter values investigated in the six experiments
  presented here as well as the symbol used to represent their
  results in figure~\ref{gc_res}.}
\label{expt_dets}
\end{table}

\begin{figure}
\centering
\includegraphics[width=10cm]{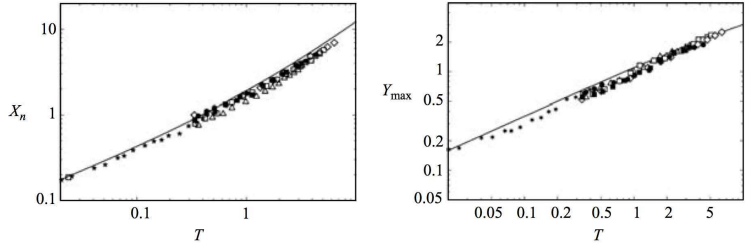}
\caption{Numerical (solid line) and experimental (points) results for
  the position of the nose of the current, $X_n$,
and the maximum
  horizontal extent of the current, $Y_{\mathrm{max}}$, as functions
  of time. The symbols used to represent each experimental run are
  given in table~\ref{expt_dets}.}
\label{gc_res}
\end{figure}

\begin{figure}
\centering
\includegraphics[height=3cm]{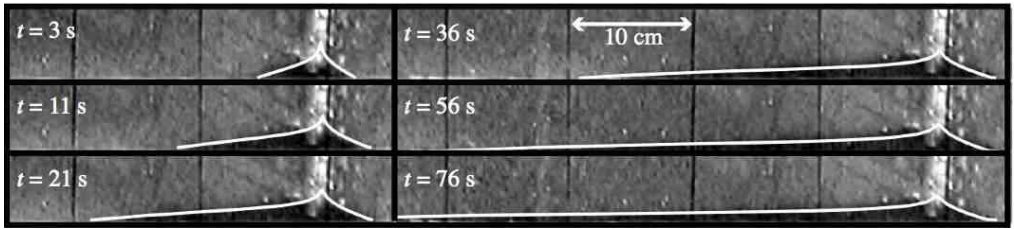}
\caption{Comparison between the current profiles (dark) observed in Experiment 3 and those predicted by numerical computation (super-imposed white lines).}
\label{profiles}
\end{figure}

The experimental results plotted in figure~\ref{gc_res} shows that the
experimental results are in good agreement with the theoretical
results produced by solving \eqref{ndpde}. The comparison between experimentally observed current profiles and those predicted from theoretical solutions of \eqref{ndpde} shown in figure \ref{profiles} is also favourable --- particularly away from the source region. Two possible mechanisms may account for the slight discrepancy between experiments and theory observed: the drag exerted by the solid substrate on the current and the fact that the pore Reynolds number in our experiments is typically $O(5)$. Such a value of the pore Reynolds number suggests that we may be approaching the regime where Darcy's law begins to break down, which is around $\textrm{Re}=10$~\cite[][]{bear88}.

\section{Geological relevance}

Our experimental and numerical analyses have shown that shortly after
the initiation of a constant flux gravity current ($\alpha=1$) it begins to spread axisymmetrically in the
manner described by~\cite{lyle05}. However, at times much longer than
the characteristic time $t^*$ given in \eqref{scales}, the current
loses its axisymmetry and propagates predominantly downslope. Since it
propagates at constant velocity in this regime, the current propagates 
much further and faster in this case than would be the case if it remained
axisymmetric. This is potentially problematic in a range of practical
applications, such as the sequestration of carbon dioxide in which
super-critical carbon dioxide is pumped into aquifers. Since the
density of the liquid carbon dioxide lies in the range
$500\pm150\mathrm{~kgm}^{-3}$~\cite[][]{chadwick05}, it remains buoyant with
respect to the ambient water and so will rise up any inclined
boundaries.

The time-scale, $t^*$, over which asymmetric spreading develops is of
interest to those wishing to predict the course of the released
current. While it is difficult to evaluate $t^*$ in a precise manner because of the
uncertainties in the properties of the surrounding rock, we can
perform some estimates on the basis of the available data from the
Sleipner field~\cite[][]{bickle05,chadwick05}. In this Norwegian field, around
$10^9\mathrm{~kg}$ of liquid $\mathrm{CO}_2$ is currently pumped into
the local sandstone each year. Presumably due to geological complications, this single input flux is observed later to separate into around ten independent currents propagating within different horizons of the permeable layer, each of which has a volume flux lying in the region $0.002 \lesssim q\lesssim 0.03\mathrm{~m}^3\mathrm{s}^{-1}$. Combined with typical
measured values for the porosity and permeability of
$0.7 \leq k \leq 5 \times 10^{-12}\mathrm{m}^2$ and $\phi=0.31\pm0.04$ as well as the $\mathrm{CO}_2$ viscosity, $\mu=3.5\pm0.5\times 10^{-5}\mathrm{~Pas}$~\cite[][]{bickle05} we can estimate upper and lower bounds on the value of $t^*$. When $\theta=1^\circ$, we find that $0.03\leq t^*\leq 14.2\mathrm{~years}$. This suggests that the effects of non-axisymmetric
spreading may indeed be important in the field. Because of the variety of values of the slope that we might
expect to encounter in any geological setting, we note also that for
$\theta\ll1$, $t^*\sim \theta^{-4/(3-\alpha)}$. For constant pumping
rate ($\alpha=1$), this gives $t^*\sim\theta^{-2}$: i.e.~the precise
value of the timescale over which the current becomes asymmetric
depends sensitively on $\theta$. This suggests that the different
spreading regimes discussed here may be observed in the field and may also have
practical implications.

Since injection occurs into confined layers of sediment, estimates for
the vertical scale of the current, $h^*$, are also important. Interestingly, $h^*$ is independent of $\theta$ for
$\theta\ll1$ (measured in radians) and $\alpha=1$ so that, with the parameter values given above, we find
$1.2 \leq h^*\leq 25~{\mathrm m}$. This suggests that, near the source, the
depth of the sediment layer may be similar to that of the current (and
so exchange, confined flows may become significant). However, we expect that the scaling
$H\sim T^{-1/3}$ valid away from the source ensures that the present
study will remain valid downstream.

\begin{acknowledgements}
We are grateful to John Lister for access to his code for a viscous
current on a slope and to Robert Whittaker for discussions. Mike Bickle, Andy Chadwick, Paul Linden and John Lister also provided valuable feedback on an earlier
draft of this paper. 
\end{acknowledgements}

\end{document}